*This manuscript has been accepted for publication and is in press in EPIDEMIOLOGY*

# The ideal trial: defining causal estimands that balance relevance and feasibility in target trial emulations and actual randomized trials

Margarita Moreno-Betancur[1,2,*], Rushani Wijesuriya[1,2], John B. Carlin[1,2]

## Abstract

Causal inference is the goal of randomized trials and many observational studies. The first step in a formal causal inference framework is to define the causal estimand, and in both types of study this can be intuitively defined as the effect in an ideal trial: a hypothetical perfect randomized experiment (with representative sample, perfect adherence, etc.). The target trial framework is increasingly used for causal inference in observational studies, but clarity is lacking in how a target trial should be specified and how it relates to an ideal trial. In this paper, we review the concept of the ideal trial and highlight the need to balance relevance for decision-making in the real world and feasibility of estimation when specifying it. We then consider the question of how a target trial should be specified, outlining the challenges of a recommended approach, commonly seen in applications, that puts the focus heavily on feasibility of estimation: to specify the target trial such that it is closely aligned with the observational data (e.g. uses the same eligibility criteria). We argue that with this “aligned” approach, biases may remain relative to the estimand of ultimate practical interest, defined by the ideal trial, which mirror the often-overlooked biases of actual trials. We conclude that consideration of the ideal trial and of how the target trial and its emulation or the actual trial differ from it is necessary to identify and manage all bias sources in both settings. An example from respiratory epidemiology is used for illustration.

[1] Clinical Epidemiology and Biostatistics Unit, Department of Paediatrics, University of Melbourne, Melbourne, Australia
[2] Clinical Epidemiology and Biostatistics Unit, Murdoch Children’s Research Institute, Melbourne, Australia
[*] Corresponding author: margarita.moreno@unimelb.edu.au

## Introduction

Recent decades have seen substantial methodological developments for causal inference, [1–5] which we understand as the task of addressing "what if" questions about the impact of interventions.[1] This is the goal of randomized controlled trials and of many observational studies. A key feature of a formal causal inference framework is the distinction between three essential steps: (1) (non-parametric) causal estimand definition, (2) (non-parametric) identification and (3) estimation (Figure 1). The causal estimand represents the quantity that investigators seek to estimate and thus the referent for assessing all potential sources of bias, whether in a causal observational study or an *actual* trial (i.e. a randomized controlled trial in the real world, which is inevitably imperfect).

One of the main contributions of the causal inference literature has been to formally define causal estimands. The canonical example is an average causal effect, which is defined as a contrast (e.g. difference, ratio) between average outcomes in the target population if everyone received an intervention versus if everyone received an alternative intervention. Such estimands, as well as other causal estimands, can be expressed mathematically using potential outcomes, which refer to outcomes that would be observed under different interventions. As we explain in detail further below, an alternative, equivalent way of defining a causal estimand is as the effect that would be obtained in an *ideal trial*[1,6]: a hypothetical randomized trial addressing the causal question of interest with a perfect design (e.g. a sample representative of the target population, perfect adherence to the assigned intervention, etc.).

Recently, there has been increased uptake of the so-called "target trial framework" [6–9] for guiding the design of causal analyses in observational studies. This approach consists of specifying the protocol of the *target trial*, a hypothetical randomized trial examining the causal question of interest, and then performing the analysis of the observational data in a way that emulates the target trial as closely as possible. The popularity of the approach may be explained by its intuitive appeal, given researchers' familiarity with randomized trials, and their longstanding status as a "gold standard" for intervention evaluation, yet there seems to be much variability in how it is understood.[10–15] This can result in potentially consequential differences in how the approach is taught, implemented, and interpreted in practice.

A key aspect on which clarity has been lacking is how a target trial should be specified and, crucially, how it relates to the ideal trial that defines the causal estimand. There has been scant discussion about this question beyond recommendations that the specified target trial needs to be pragmatic[8] (e.g. not placebo-controlled or blinded) so that its emulation is feasible with observational data. In principle, a target trial can be specified in several ways, with features that are close to those of the ideal trial, close to those of the observational data, or anywhere in between (Figure 2).

In a foundational target trial paper[6], Hernán and Robins indicated that one should specify the target trial as the hypothetical trial that is both "reasonably supported" by the observational data and closest to the ideal trial and then describe how the target trial differs from the ideal trial. However, in most discussions (e.g. a manuscript published by Hernán and colleagues [20] whilst this paper was in pre-print) and applications (e.g. [16–19]) of the approach, the ideal trial is not explicitly mentioned or considered as an important reference for guiding the target trial specification. Rather, the approach that is implicitly or explicitly suggested or employed is to specify the target trial solely with consideration for the extent to which it is reasonably supported by the observational data. Specifically, in what we refer to as the "aligned" approach, the target trial is specified so that it is closely aligned with features of the observational data (Figure 2). For example, in evaluations of pharmacological interventions using healthcare databases, the eligibility criteria may include being enrolled in the health care system to which the database corresponds, and the outcome measure in the target trial may be specified to be the same as that captured in the database.

The rationale for the aligned approach is that it narrows the range of assumptions beyond no residual confounding that are needed for unbiased estimation of the estimand defined by the target trial. However, the approach may be infeasible, particularly in studies of complex social, behavioral, and environmental exposures, which are important settings where the target trial framework is already used or recommended (e.g.[21–25]). This can occur even if the data are drawn from high-quality longitudinal cohort studies or other studies with primary data collection. More importantly, the estimands defined by an aligned target trial and the ideal trial may be very different. Thus, in losing sight of the ideal trial, researchers may lose sight of biases beyond baseline confounding, mirroring the often-overlooked biases of actual trials.

The aims of this paper are two-fold. First, we seek to review and revive the concept of the ideal trial as an alternative route to causal estimand definition, as well as to highlight the balance between relevance and feasibility that must be struck when translating a research question into a causal estimand. Second, we build on these considerations to shed light on the question of how a target trial should be specified, focusing on the potential challenges of the aligned approach.

The paper is organized as follows. We begin by introducing an example in respiratory epidemiology that will be used for illustration throughout. We then tackle the question of how a causal estimand is defined, reviewing both the mathematical and ideal trial approaches. We then consider the question of how a target trial should be specified, followed by a brief consideration of how the concepts discussed apply to actual randomized trials. We end with a summary and discussion.

**Motivating example**

The example is inspired by a published investigation of the effect of breastfeeding on risk of asthma at age 6 years using data from HealthNuts, a prospective longitudinal cohort study.[26] One of the study's specific research questions concerned the impact on asthma risk of following the World Health Organisation (WHO) guideline that "Infants should be exclusively breastfed for the first six months of life to achieve optimal growth, development and health". [27] Such a question can only be addressed using observational data because it would be unethical to randomize infants to being exclusively breastfed or not.[28] The published manuscript did not use the target trial framework, but in Table 1 we have outlined an ideal trial, a (partially aligned) target trial, and an emulation strategy with the HealthNuts data, which we will refer to throughout the manuscript to illustrate concepts and issues.

## The ideal trial defining the causal estimand

In a formal causal inference framework (Figure 1), the first step is to define the causal estimand, that is, the quantity to be estimated. Usually, this is defined mathematically, for example using potential outcomes, as we now review to motivate the specification of an ideal trial as an alternative approach.

### *Defining the causal estimand mathematically*

We introduce some notation. For a given child let $\bar{A}$ denote a vector of binary 1/0 indicators of exclusive breastfeeding each day, say, over the first 6 months of the child's life (for each day, 1 indicates exclusive breastfeeding, 0 otherwise), such that $\bar{A} = \bar{1}$ indicates perfect adherence to the WHO guideline and $\bar{A} \neq \bar{1}$ denotes a deviation from it (i.e. non-exclusive or no breastfeeding during the first six months). Let $Y$ denote the binary 0/1 outcome such that $Y = 1$ if the child has asthma at age 6 and $Y = 0$ if not and let $Y^{\bar{A}=\bar{a}}$ denote the potential outcome when $\bar{A}$ is set to $\bar{a}$.

In time-varying treatment settings, there is no single average causal effect (ACE) as there are many possible longitudinal treatment strategies that can be compared. In our example, an ACE of interest focuses on comparing the impact of following ($\bar{A} = \bar{1}$) versus not following ($\bar{A} \neq \bar{1}$) the WHO guideline, where the latter includes any deviation from the guideline, no matter its extent. Note that full adherence ($\bar{A} = \bar{1}$) may be impossible to enforce in the real world where mothers may become unable to breastfeed due to e.g. illness. Such allowed deviations could be incorporated into the treatment strategy to make it more practically relevant, however for brevity we focus here on the simple $\bar{A} = \bar{1}$ vs $\bar{A} \neq \bar{1}$ contrast. This estimand can be defined in the risk difference scale as follows:

$$\text{ACE} = \text{E}\left(Y^{\bar{A}=\bar{1}}\right) - \text{E}\left(Y^{\bar{A}\neq\bar{1}}\right). \quad (1)$$

This is interpreted as the difference in asthma risk in a thought experiment where all versus no infants in the target population are exclusively breastfed for the first 6 months of life.

The potential outcomes notation enables the mathematical representation of such thought experiments, where individuals are subject to interventions. However, several characteristics of the thought experiment remain implicit in (1). Firstly, the expectation is taken over the distribution of effect modifiers in a target population, but this is not explicit in the notation. Secondly, it is implicit that there is no outcome missingness (e.g. no loss to follow-up). Finally, it is implicit that the outcome $Y$ is perfectly measured or, perhaps more precisely, that $Y$ corresponds to the measure in the real world that most accurately operationalizes what most people would understand by "asthma", i.e. a gold standard measure.

To make these aspects explicit, we would first need to explicitly define the target population and gold standard outcome measure, as well as confirm that there is a well-defined study plan that could avoid outcome missingness in the real world. In the context of the example, although this may be subject to debate (see section "Balancing relevance vs feasibility when defining the causal estimand" below), we could define the target population as the population of neonates born in Melbourne who can be breastfed, and the gold standard outcome measure could be defined as clinical diagnosis of asthma. Outcome missingness should be avoidable by a well-defined plan, noting that the occurrence of missing outcomes due to death would be negligible given the very low infant mortality rates in Australia (<4/1000 live births after 2010 [29]).

To formally cover all these aspects, we expand and clarify the notation. Let $E_F$ denote the expectation over a distribution $F$ for the vector of effect modifiers, and let $F_1$ denote the distribution of effect modifiers in the target population. Redefine $Y$ to denote asthma as measured using clinical diagnosis. Let $M_Y$ denote a binary 0/1 outcome missingness indicator, such that $M_Y = 1$ if outcome is missing and $M_Y = 0$ otherwise. Then the ACE as described in the abovementioned thought experiment can be defined more precisely as

$$\text{ACE} = E_{F_1}\left(Y^{\bar{A}=\bar{1},M_Y=0}\right) - E_{F_1}\left(Y^{\bar{A}\neq\bar{1},M_Y=0}\right). \qquad (2)$$

***Defining the causal estimand by specifying an ideal trial***

An alternative, but equivalent and potentially more accessible way of defining the causal estimand is by directly specifying the protocol of the thought experiment that the potential outcomes notation seeks to represent. This thought experiment can be conceived as a hypothetical randomized trial under which individuals in the target population are randomized to the two interventions of interest (and they adhere to the assigned intervention). Indeed, if we imagine an infinite sample size (i.e. under asymptotic conditions), this is mathematically equivalent to contrasting the outcome risks in the whole target population under one versus the other intervention.

Specifically, we can define the causal estimand by specifying the protocol of an *ideal trial*: a hypothetical randomized experiment that would address the specific research question of interest with a perfect design. That is, in addition to randomization, it has a sample representative of the target population, perfect adherence to the assigned intervention, no missing data and no measurement error.

To illustrate, Table 1 outlines the protocol for an ideal trial in the asthma example, defining an estimand that is equivalent to the ACE in (2). That is, asymptotically, components A to E of the ideal trial fully operationalize this estimand. To see this, let $Z$ denote the randomized treatment assignment indicator in the ideal trial. Then, considering the protocol of the ideal trial, the intention-to-treat (ITT) effect and the per-protocol effect (PPE), as usually defined, are respectively given by:

$$\mathrm{ITT}_{\mathrm{ideal}} \coloneqq E_{F_1}(Y^{Z=1}) - E_{F_1}(Y^{Z=0}) = E_{F_1}(Y^{Z=1,M_Y=0}) - E_{F_1}(Y^{Z=0,M_Y=0})$$

and

$$\mathrm{PPE}_{\mathrm{ideal}} := E_{F_1}\left(Y^{\bar{A}=\bar{1}}\right) - E_{F_1}\left(Y^{\bar{A}\neq\bar{1}}\right) = E_{F_1}\left(Y^{\bar{A}=\bar{1},M_Y=0}\right) - E_{F_1}\left(Y^{\bar{A}\neq\bar{1},M_Y=0}\right)$$

where the second equality in each row is trivial because there is no outcome missingness in the ideal trial. Therefore, $\mathrm{PPE}_{\mathrm{ideal}}$ is equal to the ultimate causal estimand of interest (2). Additionally, $\mathrm{PPE}_{\mathrm{ideal}} = \mathrm{ITT}_{\mathrm{ideal}}$ because there is perfect adherence.

The approach of specifying an ideal trial has two advantages over the mathematical approach. First, all the ingredients needed to precisely define the estimand must be made explicit, as they correspond to characteristics of the thought experiment that need to be specified. That is, this approach ensures a precise definition as in (2). Second, as has been noted[6], this approach may be more accessible to those less mathematically inclined, especially given the familiarity of researchers with randomized trials.

### *Balancing relevance vs feasibility when defining the causal estimand*

Whether using the mathematical or ideal trial approach, there is rarely an a priori uniquely defined estimand for a given causal question. As described earlier for the asthma example, the process of precisely defining a causal estimand requires making specific choices about the underlying thought experiment. These choices need to strike a balance between the following two conflicting aspects:

A. Relevance to decision-making in the real world.
B. Feasibility of unbiased estimation with the available data.

The conflict arises because, generally, the estimation of an estimand that is more relevant to inform real-world decision-making will require stronger identifiability assumptions, that is, it will be less feasible to estimate it with the available data. When these assumptions are highly implausible, we may need to change the question or consider an alternative data source.

We sought to strike this balance when defining the estimand for the asthma example. For instance, consider the specification of the gold standard outcome measure. There is no single test to confirm "asthma", so clinical diagnosis, which involves medical history review, physical examination and lung function tests, is the measure that most accurately operationalizes how the condition is detected in the real world. Now, the HealthNuts study only had a parent-reported measure of asthma, based on parental reports of doctor-diagnosed asthma and use of common asthma medication. Therefore, unbiased estimation of the causal estimand that we defined would require some assumptions about the measurement error structure [30] (e.g. no unmeasured common causes of parent-reported asthma and breastfeeding practices; see also Table 2). We could have eliminated the need for assumptions by redefining the outcome to be based on the parental-reported asthma measure but doing so would have substantially reduced the real-world relevance of the estimand.

As an example of a scenario where we would need to change the question, suppose that the HealthNuts study had not collected data on asthma but had collected data on eczema, a related condition. The assumptions required to use any eczema measure as a proxy for (clinically diagnosed) asthma would be highly implausible, so we would need to either change the question to examine the impacts of exclusive breastfeeding on eczema (and define a corresponding estimand) or consider an alternative data source.

It is worth noting that these decisions surrounding the estimand specification entail important value judgements about the question that we seek to address, which can be referred to as estimand assumptions.[31] For example, balancing relevance and feasibility, we have defined the target population as comprising all neonates born in Melbourne who can be breastfed. We did not include "parent/guardian can read and understand English" as an eligibility criterion because, realistically, the findings from this study will be used to provide guidance to all families in Melbourne with neonates who can be breastfed, regardless of the parents' language. Thus, by including neonates with parents who do not read or understand English in the target population, we must explicitly consider these families and the additional assumptions required in the interpretation of findings. For similar reasons, we did not include "born in 2006-2010" in the eligibility criteria either, so researchers must consider whether findings from this dataset would apply to neonates born now, who will be those impacted by the results of the study.

**The target trial framework**

***How should the target trial be specified?***

In principle, the target trial could always be specified as the ideal trial (Figure 1). An alternative approach that is implicit or explicit in applications and discussions of the framework in the literature is to specify the target trial in a way that is closely aligned with the observational data. This includes close alignment in eligibility criteria, in occurrence of non-adherence and outcome missingness, and in outcome and adherence measures. We refer to this as the "aligned" approach to target trial specification. There are several examples of its application (e.g. [16–19]), and it is for instance the approach described in a recent paper by Hernán and colleagues, who state that the target trial is "a randomized trial that can be reasonably emulated with the available observational data". [20]

The aligned approach, which focuses heavily on the "feasibility" criterion above, has the advantage of ensuring that all protocol components apart from randomization can be closely emulated. That is, it minimizes the assumptions needed for unbiased estimation of the estimand defined by the target trial. However, the approach raises two challenges. Firstly, it may not be possible to implement in many cases. Secondly, and perhaps most critically, in setting aside the "relevance" criterion it defines an estimand that may not be very useful for informing real-world decision-making, obscuring the assumptions that would be needed to bridge the gap to practice. Next we illustrate these two challenges.

***Challenges with the aligned approach***

In the asthma example, we are not able to specify a target trial that is closely aligned with the HealthNuts data – there is necessary misalignment in the *Eligibility criteria* (columns 3 and 4 of Table 1). Specifically, a target trial would necessarily follow infants from birth but in the cohort study infants are only recruited at 11-15 months (from immunization clinics) and followed up from then. Furthermore, to be in the emulation study sample, families would have had to consent to participate in HealthNuts (a multi-faceted long-term cohort study), whereas in a target trial they would have had to consent to participate in a randomized experiment. Finally, in the emulation, the analytic sample needs to be determined with consideration for missing data in variables such as the treatment (and adherence to it), confounders and eligibility criteria, but there would be no such missing data in a target trial. Such misalignments are consequential in that they represent potential sources of causal bias beyond baseline confounding that could arise in the emulation relative to the estimand defined by the target trial. For example, if there are unmeasured common causes of reaching age 11-15 months and consenting to participate in the cohort at this post-treatment time-point, the emulation could be subject to selection bias. Multivariable missing data are also an important source of bias depending on the causal structure[32–36].

A more critical challenge is that the (partially) aligned target trial defines an estimand that is of questionable relevance to real-world decision-making. As mentioned earlier, findings from the study would realistically be used to guide all families in Melbourne with neonates who can be breastfed, not just those whose parents can read and understand English or who are willing to participate in a randomized trial. Further, selecting a measure of asthma that is based on parent report distances the estimand from the measure that is most important to families in the real world (clinical diagnosis). Additionally, decision makers are interested in the effect of following the WHO guideline under perfect adherence and absence of outcome missingness.

In other words, the (partially) aligned target trial defines an estimand that differs from the causal estimand of ultimate interest, as defined in (2) or via the ideal trial in Table 1. To show this formally, let $F_2$ denote the distribution of effect modifiers in the target population defined by the eligibility criteria in the partially aligned target trial, and $Y_*$ and $\overline{A_*}$ denote the corresponding outcome and adherence measures, then the ITT and PPE estimands in the partially aligned target trial are:

$$\mathrm{ITT}_{\mathrm{target}} := E_{F_2}(Y_*^{Z=1}) - E_{F_2}(Y_*^{Z=0})$$

and

$$\mathrm{PPE}_{\mathrm{target}} := E_{F_2}\left(Y_*^{\overline{A_*}=\overline{1}}\right) - E_{F_2}\left(Y_*^{\overline{A_*}\neq\overline{1}}\right).$$

Here, non-adherence means that $\mathrm{PPE}_{\mathrm{target}} \neq \mathrm{ITT}_{\mathrm{target}}$ in general. Further, given the discrepancies between $F_1$ and $F_2$, outcome and adherence measures and outcome missingness (note lack of superscript "$M_Y = 0$"), both these contrasts are generally different from the causal estimand of ultimate interest defined in (2) (although we note that Hernán and colleagues[20] state that the estimand implicitly includes no loss to follow-up, and thus no outcome missingness).

It follows that a study specifying the target trial using the aligned approach is subject to several potential sources of bias relative to the ultimately desired estimand. Specifically, in *Eligibility criteria,* selection bias may arise if the observational study sample is not representative of the true target population – so called type II selection or generalizability bias[37]. In the example, this could arise if the effect of breastfeeding on asthma differs by whether the parent can read and understand English, which is possible if there is effect modification by ethnicity, for which English fluency is a strong proxy. Further selection bias may arise due to missing data in the outcome, or in predictors of either missingness in outcome, selection into the study sample or adherence to treatment. For instance, bias can arise if there are unmeasured common causes of outcome and outcome missingness such as parental healthcare-seeking behavior. In *Treatment strategies,* confounding bias could arise in estimation of the per-protocol effect if there are unmeasured predictors of non-adherence and outcome, and measurement bias could arise due to error in the adherence measure based on retrospective parent report. In *Outcome*, measurement bias could arise from the target trial relying on the parent-reported asthma measure rather than clinical diagnosis.

***Implications for practice***

In an observational study, a set of causal assumptions are required to identify the causal estimand of ultimate interest (2), whether specified mathematically or via an ideal trial. These assumptions are independent of how the target trial is specified on the spectrum from ideal to closely aligned with the observational study (Figure 2): indeed, the target trial specification only affects the assumptions required for identifying the ideal trial estimand in the target trial itself.

Therefore, if in practice the target trial is specified to be identical to the ideal trial, then no assumptions are needed to go from the target to the ideal trial, and all causal identification assumptions that must be considered in the analysis and interpretation pertain to those needed to connect the target/ideal trial with the emulation with observational data (Figure 1).

If instead the target trial specification deviates from the ideal trial – as we showed is generally the case when using the aligned approach – the causal identification assumptions that researchers need to consider can be partitioned into two sets: those required to identify the ideal trial estimand within the target trial, and those required to emulate the target trial with the observational data (Figure 1). In this setting, the former assumptions (and thus associated biases) remain tacit unless the ideal trial is specified. By this reasoning, the recommendation from a foundational target trial paper[6] remains key: researchers should specify the ideal trial, explain how the target trial deviates from it and assess the additional assumptions implied.

**Actual randomized trials**

Actual randomized trials must also begin by defining the causal estimand that they target, either mathematically or by specifying an ideal trial. Although not subject to baseline confounding, such a study generally deviates from the ideal trial in that it is subject to non-generalizability, non-adherence and missing outcome data. Thus, causal identification assumptions about these processes are needed for identifying the estimand defined by the ideal trial (Figure 1), and if these are violated, causal biases relative to this estimand would arise. Therefore, the interpretation of findings from actual randomized trials should also consider the ideal trial and explain how the actual trial deviates from it. Of note, in cases where an actual trial explicitly aims to estimate the effect of allocation rather than adherence to the randomized interventions, then in the ideal trial the treatment strategies would be specified as allocation to rather than receipt of the interventions.

**Summary: ideal, target and actual trials**

Figure 1 summarizes how ideal trials, target trials and actual trials fit within a formal causal inference framework[1–5], and delineates the assumptions involved in each stage. The ideal trial defines the causal estimand of ultimate interest, whether the study is an observational study emulating a target trial or an actual trial.

Table 2 details the issues that may make an actual trial and a target trial emulation in an observational study different from the ideal trial, and the corresponding causal assumptions required for identifying the estimand that the ideal trial defines. If these assumptions are violated, causal biases relative to this estimand will arise.

While the *Statistical analysis plan* for the ideal trial is trivial, that for a target trial emulation or for the analysis of an actual trial must delineate the analytic approaches that will be used in light of the causal identification assumptions, for example to adjust for non-adherence and missingness given the selected adjustment variables (e.g. selected confounders, predictors of missingness). Analytic approaches often rely on parametric modelling assumptions. The execution of this plan when analyzing the data corresponds to the estimation step, and it is at this stage that one must reckon with finite sample size issues and potential non-causal biases (e.g. due to misspecification of parametric models).

**Discussion**

Given the accelerating uptake of the target trial framework over recent years, bringing to light nuances in its implementation and interpretation is important for strengthening its use in practice. In this paper, we considered the question of how a target trial should be specified and highlighted challenges with the increasingly popular "aligned" approach. The main issue is that this approach puts the focus heavily on the feasibility of unbiased estimation with the available data, resulting in an estimand that may not be relevant to addressing the question of most interest to real-world decision-making.

We argued that instead (or in addition), researchers should consider the ideal trial that would be specified by starting with the question of interest and then refining it by seeking a balance between relevance and feasibility. As such, the ideal trial brings to the fore assumptions that may be otherwise overlooked when seeking to bridge the gap between the study findings and how they will be used in practice. Understanding these assumptions is important to avoid an overconfidence in evidence arising from target trial emulations that mirrors the generally acknowledged overconfidence in evidence arising from actual trials[38,39]. For instance, it has been suggested that a correct target trial emulation eliminates common sources of bias except possibly for baseline confounding,[9] but this is relative to the specified target trial and the study may still be subject to other important biases relative to the ideal trial. A key source of this overconfidence is an overemphasis on internal validity, with issues like non-representativeness of the sample being relegated to the secondary plane of external validity, with no attempt to address them. Westreich et al.[39] have elegantly argued the need to move away from the internal vs external validity distinction and focus on overall (so-called "target") validity, and the ideal trial provides the reference for examining that.

We have illustrated how precisely defining the causal estimand, whether mathematically or via the specification of an ideal trial, entails a tricky balancing act and important value judgements. However, going through this process may not only improve but facilitate the process of analysis design and interpretation of findings. In particular, although researchers invariably consider the limitations of their study in their interpretation, the process by which they identify and list these limitations (e.g. in the discussion of a manuscript) can be haphazard: there is no "checklist", which can lead to selective consideration and reporting. As we have shown elsewhere [40], using the structured approach of specifying the ideal trial as an additional column in a table, and contrasting it with the emulation, serves essentially as a checklist that facilitates the systematic identification of all potential sources of bias, and thus their comprehensive reporting and consideration in study interpretation.

We have argued that specifying the ideal trial is important for actual trials too and it is also fundamental in causal investigations that bring together data from multiple studies: multiple trials, multiple observational studies[41] or combinations of both, such as when benchmarking observational studies against an actual trial[42–44]. Indeed, comparing one study with another is the usual practice, whereas in the logic of causal inference, each study should be appraised against a common reference, defined in terms of the causal estimand of ultimate interest (i.e. the ideal trial), as has been argued elsewhere.[41] Taking this approach, a study would present a table with one column for the ideal trial specification, and then potentially several columns, one for each emulation of it with data from observational datasets or actual trials. [41] This is similar in spirit to the harmonized protocol presented by Lodi et al. [42], who compared an actual trial and an observational study, but their reference point was an aligned rather than an ideal protocol.

Of note, there have been other instances of additional columns in studies applying the target trial framework, although they have been in a somewhat different spirit to what we propose here. For example, some studies (e.g. García-Albéniz et al.[45]) use an actual trial, termed the "index trial", to motivate the specification of the target trial that they then emulate using observational data. Of course, as we have argued, an actual trial is inevitably imperfect, so this is quite distinct to specifying an ideal trial.

To conclude, biases may remain relative to the estimand of ultimate practical interest when using the "aligned" approach to specify the target trial. Consideration of the ideal trial and of how the target trial and its emulation or the actual trial differ from it is necessary to delineate all potential sources of bias in both settings.

## TABLES

**Table 1. Ideal trial, partially aligned target trial, and emulation implicit in observational study analysis, for the asthma example**

| Protocol component | Ideal trial | Partially aligned target trial[a] | Emulation with HealthNuts cohort study |
|---|---|---|---|
| A. Eligibility criteria[b] | Study sample<br>• Representative sample of<br>  o neonates born in Melbourne who can be breastfed, regardless of the language spoken by their parents or guardians<br><br>Analytic sample<br>• All participants in study sample | Study sample<br>• Representative sample of<br>  o neonates born in Melbourne in 2006-2010 who can be breastfed<br>  o with a parent/guardian who can read and understand English<br>  o whose families consent to participate in a randomized experiment [f]<br><br>Analytic sample<br>• All participants in study sample regardless of missing data in outcome | Study sample<br>• Sample of<br>  o infants aged 11-15 months in 2007-2011 participating in 170 council-run immunization sessions across Melbourne who can be breastfed<br>  o with a parent/guardian who can read and understand English<br>  o whose families consent to participate in a cohort study<br><br>Analytic sample<br>• All participants in study sample regardless of missing data in exposure, outcome and confounders |
| B. Treatment strategies [c] | Treatment arms<br>• Exclusive breast-feeding during the first 6 months<br>• Non-exclusive or no breast-feeding during the first 6 months<br><br>Measure of adherence: Objective and continuous-time (day-to-day) measurement of age at initiating formula or other foods | Treatment arms<br>• Exclusive breast-feeding during the first 6 months<br>• Non-exclusive or no breast-feeding during the first 6 months<br><br>Measure of adherence: Based on 12-month parent report of age at initiating formula or other foods | Treatment arms<br>• Exclusive breast-feeding during the first 6 months<br>• Non-exclusive or no breast-feeding during the first 6 months<br><br>Measure of adherence: Based on 12-month parent report of age at initiating formula or other foods |
| C. Assignment procedures | Randomization at recruitment without blind assignment | Randomization at recruitment without blind assignment | No randomization, no blinding. Individuals classified at time zero of follow-up (i.e. birth) to strategy consistent at birth with the 12-month parent report of age at initiating formula or other foods. Individuals initiating exclusive breast-feeding at birth cannot be unambiguously classified into these strategies at that time-point. Therefore, the method of random allocation or cloning will be needed[6]. Grace periods could be used if relevant data were available. |
| D. Follow-up period | Follow-up<br>• Starts: Birth<br>• Ends: Child aged 6 years (there is no loss to follow-up) | Follow-up<br>• Starts: Birth<br>• Ends: Child aged 6 years or until loss to follow-up | Follow-up<br>• Starts: Child aged 11-15 months<br>• Ends: Child aged 6 years or until loss to follow-up |
| E. Outcome | Outcome measure<br>• Clinically diagnosed asthma at age 6, with systematic ascertainment | Outcome measure<br>• Parent-reported measure of asthma based on reports of doctor-diagnosed asthma or use of common asthma medication in the previous 12 months | Outcome measure<br>• Parent-reported measure of asthma based on reports of doctor-diagnosed asthma or use of common asthma medication in the previous 12 months |
| Causal contrasts[d] | $\mathrm{ITT}_{\mathrm{ideal}}$<br>$\mathrm{PPE}_{\mathrm{ideal}}$<br><br>Given perfect adherence, $\mathrm{ITT}_{\mathrm{ideal}}=\mathrm{PPE}_{\mathrm{ideal}}$ | $\mathrm{ITT}_{\mathrm{target}}$<br>$\mathrm{PPE}_{\mathrm{target}}$ | $\mathrm{ITT}_{\mathrm{obs}}$<br>$\mathrm{PPE}_{\mathrm{obs}}$ |
| Identifying assumptions[e] | None/trivial, as there is a representative sample, perfect adherence, no confounding, no missing data, (in particular, no loss to follow-up that would result in missing outcome), and no outcome misclassification in the ideal trial | No uncontrolled predictors of<br>- selection into study sample<br>- outcome missingness<br>- adherence at baseline and post-baseline<br>- misclassified outcome<br>that would lead to bias | No uncontrolled predictors of<br>- selection into study sample<br>- missingness in any variable<br>- adherence at baseline and post-baseline<br>- misclassified outcome<br>- misclassified exposure<br>- treatment strategy initiation (i.e. baseline confounders)<br>that would lead to bias |
| Statistical analysis plan | $\mathrm{ITT}_{\mathrm{ideal}}$, $\mathrm{PPE}_{\mathrm{ideal}}$: Compare proportions with asthma between treatment arms | $\mathrm{ITT}_{\mathrm{target}}$: Compare proportions with asthma between treatment arms, using an approach such as multiple imputation (MI) or inverse probability weighting (IPW) to handle missing outcome data due to loss to follow-up<br><br>$\mathrm{PPE}_{\mathrm{target}}$: Compare proportions with asthma between treatments arms, using MI or IPW to handle missing outcome data and IPW to adjust for non-adherence (individuals who deviate from assigned strategy are censored)<br><br>*N.B. No attempt to correct for a non-representative sample, or misclassified exposure and outcome – stronger assumptions needed* | $\mathrm{ITT}_{\mathrm{obs}}$: Cannot be emulated in this case given treatment not unambiguously assigned at birth (Hernán, Sauer et al. 2016), but if this were not the case, we would compare proportions with asthma between treatment arms, using an approach like g-computation or IPW to adjust for baseline confounders and handle missing outcome data due to loss to follow-up (MI can also be used for the latter).<br><br>$\mathrm{PPE}_{\mathrm{obs}}$: Compare proportions with asthma between treatment arms, using g-computation or IPW to adjust for baseline confounders, MI or IPW to handle multivariable missing data and IPW to adjust for non-adherence (individuals who deviate from assigned strategy are censored)<br><br>*N.B. No attempt to correct for a non-representative sample, or misclassified exposure and outcome – stronger assumptions needed* |

ITT, Intention to treat; PPE, per-protocol effect; MI, multiple imputation; IPW, inverse probability weighting.

[a] As defined in the text, this refers to partial alignment with the observational data (here, drawn from the HealthNuts cohort study). [b] Following Lu et al. 2022,[37] the target population is the population about which inference is sought, the study sample is the population enrolled into the study (trial or cohort), and the analytic sample is the portion of the study sample used in the analysis. [c] Non-adherence is defined here as crossing over to the other treatment strategy. [d] See text for formal definitions of $\mathrm{ITT}_{\mathrm{ideal}}$, $\mathrm{PPE}_{\mathrm{ideal}}$, $\mathrm{ITT}_{\mathrm{target}}$, $\mathrm{PPE}_{\mathrm{target}}$. If using breastfeeding initiation in the first day ($A_{*0}$) to operationalize treatment assignment, the observational analogs for the ITT and PPE are defined as $\mathrm{ITT}_{\mathrm{obs}} := E_{F_3}(Y_*^{A_{*0}=1}) - E_{F_3}(Y_*^{A_{*0}=0})$ and $\mathrm{PPE}_{\mathrm{obs}} := E_{F_3}(Y_*^{\overline{A}_*=\overline{1}}) - E_{F_3}(Y_*^{\overline{A}_*\neq\overline{1}})$, where $F_3$ is the distribution of effect modifiers in the emulation study sample. [e] See also Table 2. [f] Included per Hernán and colleagues[20], although we note that this contradicts the recommendation therein that only criteria that can be emulated with the observational dataset should be included in the target trial specification. Per the text, we find it unnecessary because not only is it not feasible to emulate this, but it is not practically relevant either.

**Table 2. Statistical issues that can arise in ideal trials, actual trials and target trial emulations, and full set of assumptions required for identification of the estimand defined by the ideal trial**

| Protocol component | Issue | Can the issue be present in: Ideal trial? | Actual trial? | Target trial emulation with observational data? | Assumption needed for identifying the estimand defined by the ideal trial in the presence of the issue [a] |
|---|---|---|---|---|---|
| A. Eligibility criteria | Non-representative study sample | No | Yes | Yes | No uncontrolled predictors of selection that could induce bias |
| | Missing data | No | Yes; missing data in outcome, and predictors of adherence, missingness and selection | Yes; missing data in outcome, treatment, confounders, eligibility criteria and in predictors of adherence, missingness, selection and misclassification | No uncontrolled predictors of missingness that could induce bias |
| B. Treatment strategies | Non-adherence | No | Yes | Yes | No uncontrolled predictors of non-adherence that could induce bias |
| | Exposure measurement error | No | No | Yes | No uncontrolled predictors of exposure measurement error or a form of measurement error that could induce bias |
| C. Assignment procedures | Confounding | No | No | Yes | No uncontrolled confounders at baseline that could induce bias |
| D. Follow-up period | Misalignment between time zero, treatment assignment and eligibility | No | No | Yes | [This can lead to a non-representative study sample and/or measurement error in exposure – both issues covered above] |
| E. Outcome measure | Outcome measurement error | No | Yes | Yes | No uncontrolled predictors of outcome measurement error or a form of measurement error that could induce bias |

[a] Whether uncontrolled predictors can cause bias in each case depends on the causal diagram/structure, e.g. see Zhang et al. 2024[33] for the case of multivariable missingness. For measurement error, bias depends both on the causal structure and the nature of the relationship between the construct and measured variable (e.g. if it is nonmonotonic or depends on a third variable).[30] In general, measurement error in exposure and/or outcome will cause bias except under specific conditions.[30,46]

## FIGURES

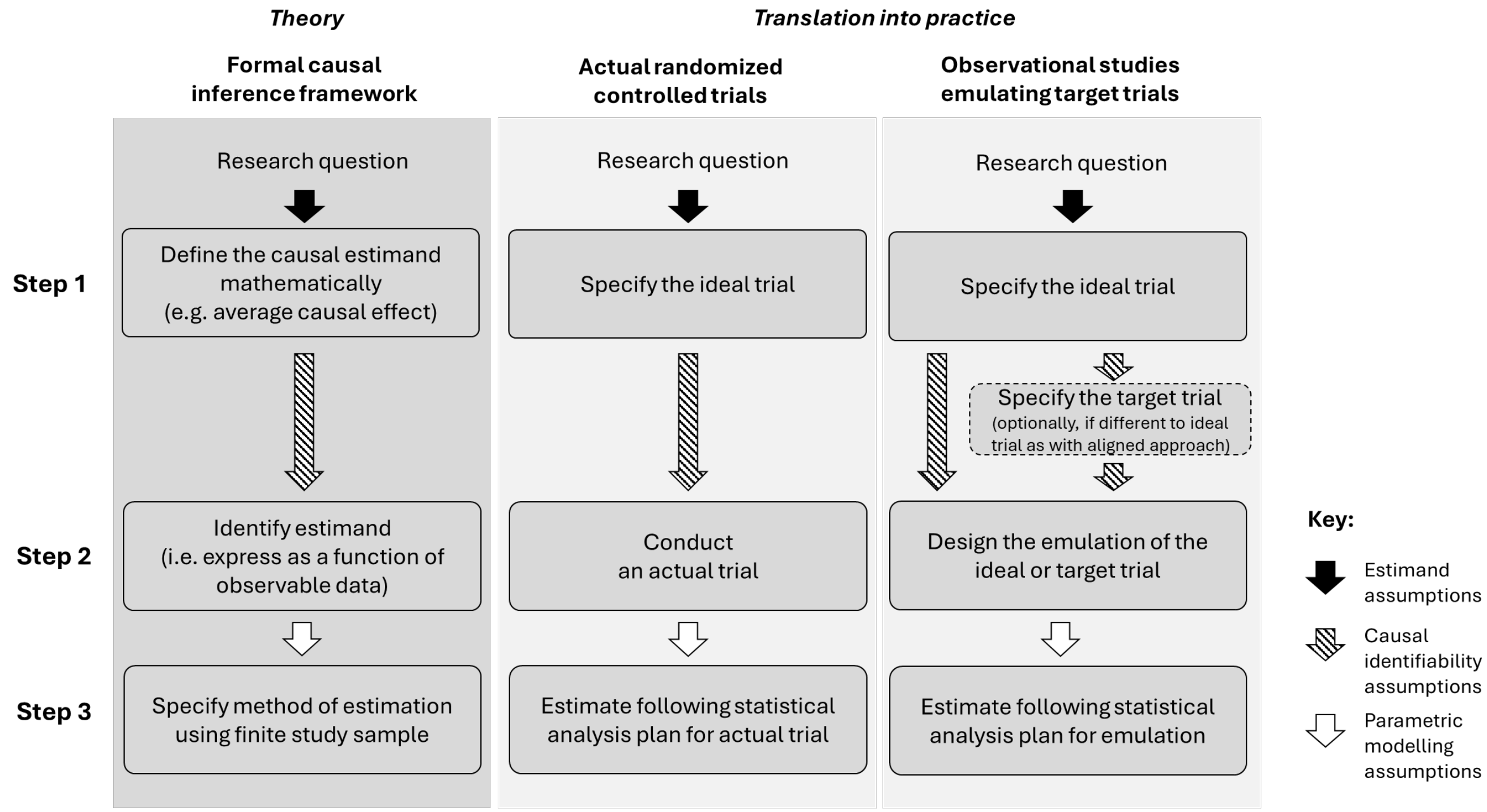

**Figure 1.** Diagram depicting how ideal, target and actual trials relate to the three essential steps in a formal causal inference framework. The target trial specification straddles Steps 1 and 2, because although the target trial serves to specify a causal estimand, if the target trial differs from the ideal trial, then as explained in the text causal identifiability assumptions are required to go from the target trial to the ideal trial.

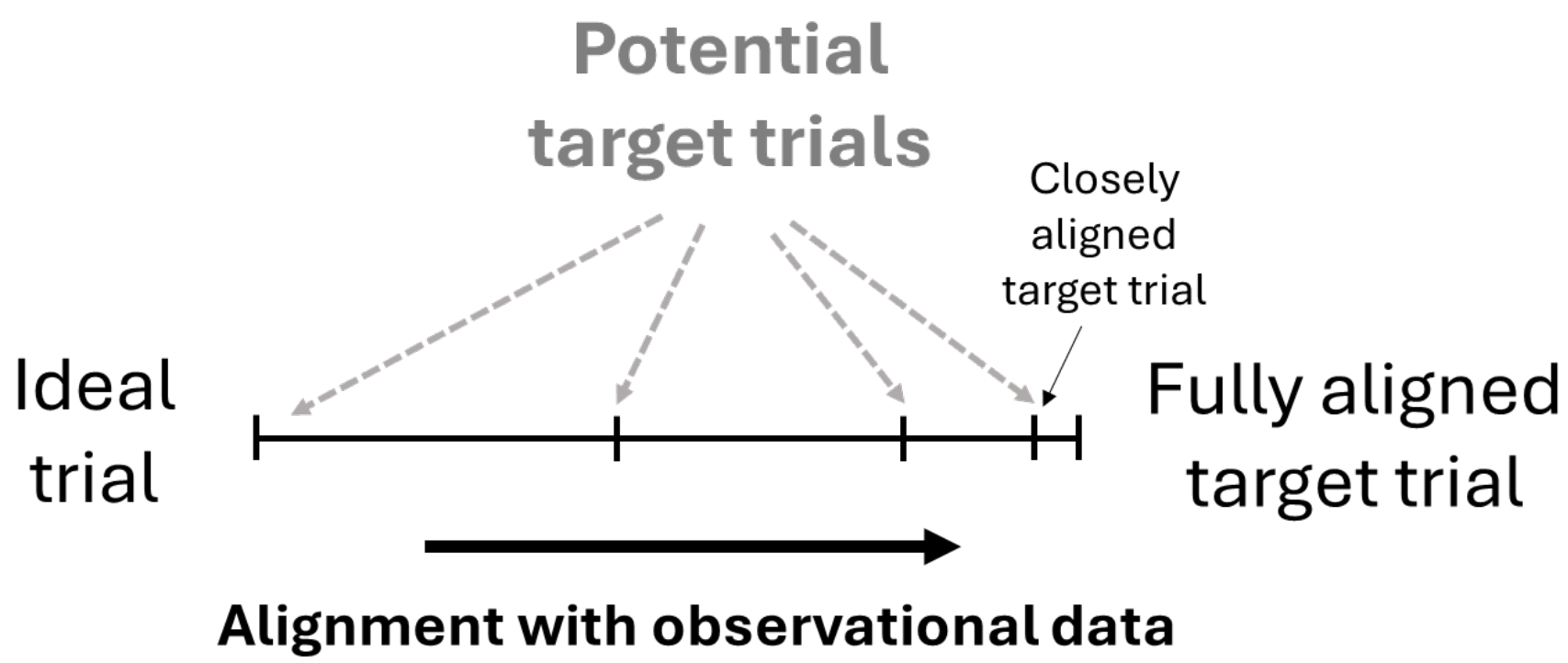

**Figure 2.** Diagram depicting how a target trial may be situated at various points on an alignment scale, from the ideal trial to closely aligned with the observational data.

**SOURCES OF FUNDING:** This work was supported by an Investigator Grant fellowship to MMB from the National Health and Medical Research Council [grant ID: 2009572]. The Murdoch Children's Research Institute is supported by the Victorian Government's Operational Infrastructure Support Program.

**CONFLICTS OF INTEREST:** None declared.

**ACKNOWLEDGEMENTS:** The authors would like to thank Rachel Peters for providing details about the analysis of the cited breastfeeding and asthma publication that we used as the basis for the example. We would also like to thank Elizabeth Stuart, Miguel Hernán and colleagues from our causal inference research team for helpful discussions and critical comments relating to this paper. Finally, we would like to thank the Editor and three reviewers for their insightful comments that helped us strengthen the paper.